\documentclass[]{spie}  %>>> use for US letter paper
\usepackage[]{graphicx}
\usepackage[]{hyperref}

\title{Wolf-Rayet stars probed by AMBER/VLTI\thanks{based on
    observations made at the ESO Paranal observatory under the run
    numbers 074.A-9025, 075.A-9001, 078.D-0503, 078.D-0656, 079.D-359
    and from archival data from the ESO Science Archive Facility.}}

\def\deg{$^\circ$}
\def\arcsec{"}

\author{Florentin Millour\supit{a},
  Olivier Chesneau\supit{b},
  Thomas Driebe\supit{a},
  Romain G. Petrov\supit{b},
  Daniel Bonneau\supit{b},
  Luc Dessart\supit{c},
  Karl-Heinz Hofmann\supit{a},
  Gerd Weigelt\supit{a},
  \skiplinehalf
  \supit{a}Max-Planck Institut for Radioastronomy, Auf dem h\"ugel, 69,
  53121, Bonn, Germany; \\
  \supit{b}Fizeau Laboratory, Nice university, Parc Valrose, Nice,
  France.\\
  \supit{c}Steward Observatory, 933 North Cherry Avenue, Tucson, AZ
  85721, USA.\\
}

\authorinfo{Further author information: (Send correspondance to
  F. Millour)\\F. Millour: E-mail: fmillour@mpifr-bonn.mpg.de,
  Telephone: +49 228 525 188}

\begin{document} 
\maketitle 

% {\bf Copyright 2008 Society of Photo-Optical Instrumentation
%  Engineers.} This paper will be published in the proceeding of SPIE
%  ``astronomical Telescopes and Instrumentation: Optical and Infrared
%  Interferometry'' and is made available as an electronic reprint
%  (preprint) with permission of SPIE. One print or electronic copy may
%  be made for personal use only. Systematic or multiple reproduction,
%  distribution to multiple locations via electronic or other means,
%  duplication of any material in this paper for a fee or for commercial
%  purposes, or modification of the content of the paper are
%  prohibited.

%%%%%%%%%%%%%%%%%%%%%%%%%%%%%%%%%%%%%%%%%%%%%%%%%%%%%%%%%%%%% 
\begin{abstract}
  Massive stars deeply influence their surroundings by their
  luminosity and the injection of kinetic energy. So far, they have
  mostly been studied with spatially {\it unresolved} observations,
  although evidence of geometrical complexity of their wind are
  numerous. Interferometry can provide spatially {\it resolved}
  observations of massive stars and their immediate vicinity. Specific
  geometries (disks, jets, latitude-dependent winds) can be probed by
  this technique.

  The first observation of a Wolf-Rayet (WR) star ($\gamma^2$ Vel)
  with the AMBER/VLTI instrument yielded to a re-evaluation of its
  distance and an improved characterization of the stellar components,
  from a very limited data-set. This motivated our team to increase the
  number of WR targets observed with AMBER. We present here new
  preliminary results that encompass several spectral types, ranging
  from early WN to evolved dusty WC.

  We present unpublished data on WR79a, a massive star probably at the
  boundary between the O and Wolf-Rayet type, evidencing some
  Wolf-Rayet broad emission lines from an optically thin wind. We also
  present new data obtained on $\gamma^2$ Vel that can be compared to
  the up-to-date interferometry-based orbital parameters from North et
  al. (2007). We discuss the presence of a wind-wind collision zone in
  the system and provide preliminary analysis suggesting the presence
  of such a structure in the data. Then, we present data obtained on 2
  dusty Wolf-Rayet stars: WR48a-b and WR118, the latter exhibiting
  some clues of a pinwheel-like structure from the visibility
  variations.

\end{abstract}

% >>>> Include a list of keywords after the abstract 


\section{Introduction}
\label{section:intro}
According to the classical MK spectral classification system, the A, B
and O stars represent the group of hot or early type stars. The most
permanent luminous stellar objects are the {\bf O stars}. They exhibit
strong, fast stellar winds of relatively low opacity, driven by their
strong radiation field. Luminosities are found to be up to $10^6
L_{\odot}$ (for O super-giants) with mass loss rates $\dot{M}$ of up to
5 $\times 10^{-6} M_{\odot}$. Terminal wind velocities ($V_{\infty})$
of up to 2000 km/s are found. Estimated surface temperatures are found
to lie up to $\sim$ 50\,000 K. In an evolutionary context, O stars are
the probable progenitors of Wolf-Rayet stars. Intense emission lines
of various ionized elements are the spectral characteristics of
Wolf-Rayet stars, originating from an extended, and rapidly expanding
atmosphere. Among all stable stars, WR stars reveal the strongest mass
loss via (radiative) wind mechanisms. Typical mass loss rates range
from $(2-10)\times 10^{-5} M_{\odot}\,yr^{-1}$, with wind velocities
of 1000 - 2500 km/s. The Wolf-Rayet winds are mostly optically thick,
so the surface is not visible. This means that the connection of
spectral classification to photospheric core temperature is not
possible as for normal stars via MK classification, and the WR
classification is purely spectroscopic. WR stars are highly luminous,
evolved, hot, massive stars, in the final state of their nuclear He
burning. They create significant quantities of heavy elements and
enrich the interstellar medium through mass-loss, and are also
candidate as progenitor of Gamma-ray bursts.

It is of importance to probe the wind of these star, as the wind is
the spectral reference for the classification and the modeling of
these extreme stars. Long baseline interferometry can help
constraining the spatial scale of the continuum and line forming
regions, only weakly constrained for these objects by classical
photometric and spectroscopic methods. % AMBER short description
AMBER (Astronomical Multi BEam Recombiner) is the VLTI (Very Large
Telescope Interferometer) beam combiner operating in the
near-infrared\cite{2007A&A...464....1P}. The instrument uses spatial
filtering with fibers\cite{2000SPIE.4006..299M}. The interferometric
beam passes through anamorphic optics compressing the beam
perpendicularly to the fringe coding in order to be injected into the
slit of a spectrograph. The instrument can operate at spectral
resolutions of 35, 1500 and 10,000, and efficiently deliver spectrally
dispersed visibilities, closure phases and differential phases.

The paper is organized following increasing evolutionary stage of the
different Wolf-Rayet stars observed so far. In
Sect.~\ref{section:WR79a}, we describe observations made on the star
WR79a, at the boundary between O and WR types. In
Sect.~\ref{section:gammavel}, we present new AMBER observations made
on the WC-type $\gamma^2$ Velorum where we probe for the presence of a
wind-wind collision zone. Then, in the last section, we present data
acquired on two dusty Wolf-Rayet stars, where the observed data points
are reminiscent of a ``pinwheel'' nebula.

\section{The ``young'' Wolf-Rayet star WR79a}
\label{section:WR79a}

HD 152408 (also known as WR79a) is probably just at the boundary
between Wolf-Rayet and O spectral types. This O-type super-giant star
shows typical broad Wolf-Rayet emission lines, indicating a higher
wind density than usual O-type stars. Therefore, this is a very nice
candidate to probe the strong changes occurring to stars from a dense
but still optically thin wind of O-type stars to a fully radiatively
driven optically thick and dense Wolf-Rayet wind.

This star was already partially successfully observed with AMBER in
2005. A reassessment of the data, with the latest version of the data
reducing software has shown a small, but significant, signal in the
Br$\gamma$ line. This star do not show a significant polarization in
the continuum, but may present a polarization enhancement in H$\alpha$
line, that may be linked to a latitude-dependent stellar
wind\cite{2002MNRAS.337..341H}. It is therefore important to use the
Br$\gamma$ line, as a probe of radial structure of the wind (density
and velocity), and to test whether it departs significantly from a
spherical symmetry.

\begin{figure}[htbp]
  \begin{center}
    \includegraphics[width=10cm]{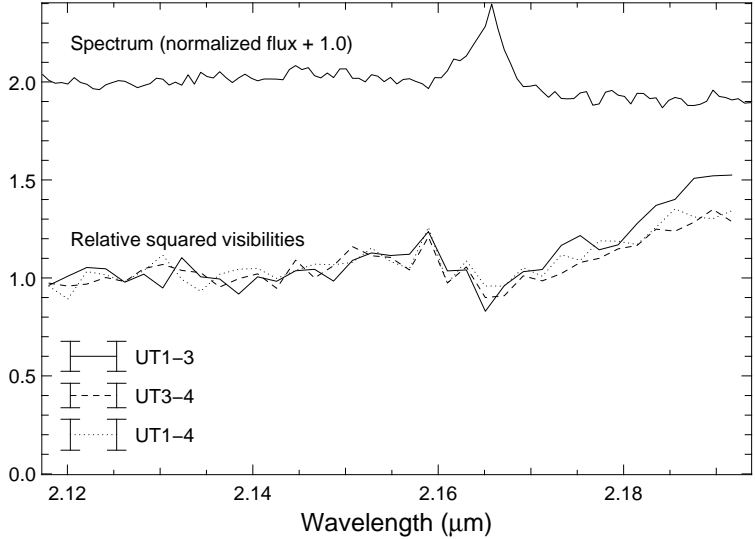}
  \end{center}
  \caption
  % >>>> use \label inside caption to get Fig. number with \ref{}
  { \label{fig:relVis} 
    The relative visibilities recorded on WR79a for 3 baselines in
    2005. {\bf A drop of visibility in the Br$\gamma$ line is seen for
      ALL baselines, related to the probably much larger spatial
      extent of the line forming region than the continuum one.}
  }
\end{figure}

AMBER observations of HD152408  were carried out during the AMBER
Science Demonstration Time in 2005 but at this time and for such
`faint' target (m$_K=4.9$), both the observing procedure and data
reduction software  were not ready to perform successful observations
in the medium resolution mode required for isolating the signal from
the Br$\gamma$ line. In the frame of recent AMBER data reduction
developments (AMBER DRS 2.0 Beta and detector fringes
removal\cite{2008A&A...479..589L}), we
have investigated again the data obtained at this time and managed to
compute relative visibilities (see Fig.~\ref{fig:relVis}). However,
due to the data noise and lack of a suitable calibration star, no
absolute visibility nor relevant differential or closure phase could
be computed.

The relative visibilities observed at this time show a clear variation
in the blended Br$\gamma$-He{\sc i} line, indicating a larger wind
envelope in this complex emission line than in the continuum, as
expected by radiative transfer modelling. However we currently cannot
extract relevant information on this envelope given the very partial
information we get from this data.

\section{The colliding-winds binary $\gamma^2$ Velorum}
\label{section:gammavel}

{$\gamma^2$~Velorum} ({WR 11}, {HD 68273}) is the closest known Wolf-Rayet
(WR) star, with an Hipparcos-determined distance of
258$^{+41}_{-31}$\,pc\cite{1997NewA....2..245V, 1997ApJ...484L.153S},
whereas other WR objects  lie at $\sim$1\,kpc or beyond.
It is also a well-studied SB2 spectroscopic binary
WR+O system (WC8+O7.5{\sc iii}, P = 78.53 d\cite{1997A&A...328..219S,
  1999A&A...345..163D}) offering access to fundamental parameters of
the WR star, usually obtained indirectly through the study of its
dense and fast wind.

This object was observed by the Narrabri intensity interferometer
operating around 0.45\,$\mu$m as early as
1968\cite{1970MNRAS.148..103H}. By observing such a star with a long
baseline interferometer, one may constrain various parameters, such as
the binary orbit, the brightness ratio of the two components, the
angular size associated with both the continuum and the lines emitted
by the WR star.

The WR component of the {$\gamma^2$~Velorum} system is of a WC8
spectral type. While 50\% of such WR stars (and 90\% of the WC9 type)
show heated ($T_d\sim1300K$) circumstellar amorphous carbon dust, ISO
observations of {$\gamma^2$~Velorum} revealed no such dust
signatures\cite{1997NewA....2..245V}. Keck observations resolved,
although only barely, the system in the $K$ band and confirmed the
absence of any dust emission from this
system\cite{2002ASPC..260..331M}, suggesting that if dust is created
near the wind-wind collision zone, it is in small amounts.

Our team observed $\gamma^2$~Velorum in December 2004 using the
AMBER/VLTI instrument with the telescopes UT2, UT3, and UT4 on
baselines ranging from 46\,m to 85\,m. It delivered spectrally
dispersed visibilities, as well as differential and closure phases,
with a resolution $R=1500$ in the spectral band 1.95-2.17\,$\mu$m. We
interpreted these data in the context of a binary system with
unresolved components, neglecting in a first approximation the
wind-wind collision zone flux contribution (Millour et al. 2007,
A\&A\cite{2007A&A...464..107M}, hereafter paper I).

Based on the accurate spectroscopic orbit and the Hipparcos distance,
the expected absolute separation and position angle at the time of
observations were 5.1$\pm$0.9mas and 66$\pm$15\deg,
respectively. However, using theoretical estimates for the spatial
extent of both continuum and line emission from each component, we
inferred a separation of 3.62$^{+0.11}_{-0.30}$\,mas and a position angle
of 73$^{+9}_{-11}$\deg. Our analysis thus implied that the binary
system lies at a distance of 368$^{+38}_{-13}$\,pc, in agreement with
recent spectro-photometric estimates, but significantly larger than the
Hipparcos value of 258$^{+41}_{-31}$\,pc. A parallel observation with
SUSI by North et al.\cite{2007MNRAS.377..415N} confirmed our astrometric point and
refined the direct distance estimate of $\gamma^2$~Velorum to be
336$^{+8}_{-7}$\,pc.

We present new AMBER data acquired with different spectral resolutions and
spectral coverage, from H to K bands. As a near-infrared signature of a wind-wind collision zone was
discussed in the paper I, we investigate in further details the
presence or not of a third light source in the system, using the full available data set acquired at five different orbital points, in order to account for, as best as possible, the binary signature of the system.

\subsection{Previous AMBER observations and new data}
\label{subsection:observations}

\begin{figure}[htbp]
  \begin{center}
    \includegraphics[width=8cm]{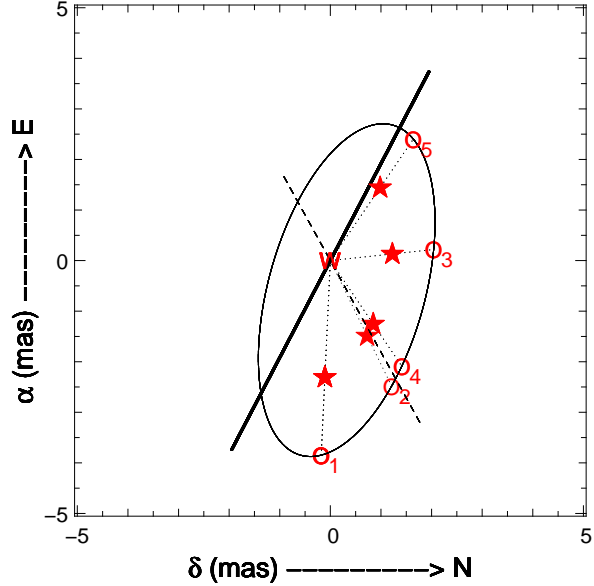}
  \end{center}
  \caption
  % >>>> use \label inside caption to get Fig. number with \ref{}
  { \label{fig:orbit} 
    Sketch of the orbit of $\gamma^2$ Velorum. The thick black line is
    the line of node, whereas the dashed line is the
    periastron-apoastron line. Over-plotted are the positions of the O
    star relative to the WR one at the different epochs of
    observation. The star symbol represent the approximate location of
    an additional light-source in the system, accounting for 5-10\% of
    the overall flux, that might be linked to the wind-wind collision
    zone.
  }
\end{figure}

% Observations parameters
{$\gamma^2$~Velorum} was observed during a follow-up campaign in 2006
and 2007 after the 2004 AMBER first scientific light, with the
initial goal to characterize the orbit of the system. During this
campaign, K-band medium spectral resolution observations were
repeated, as well as H-K low spectral resolution. We also take the
opportunity of this paper to present archival data of the very
first light of AMBER in H-band medium spectral resolution that was
made on $\gamma^2$~Velorum. The corresponding observing log is
presented in Table~\ref{obsLog} and the orbit coverage, is shown in
Figure \ref{fig:orbit}.

\begin{table}[htbp]
  \centering
  \caption{
    \footnotesize{
      Log of the observations and atmospheric conditions for
      {$\gamma^2$~Velorum}. The data-set is mainly divided in 5 epochs
      corresponding to different dates, telescopes configurations and
      spectral resolutions.
    }
  }
  \label{obsLog}

  \begin{tabular}{|c|c|c|c|c|c|c|c|c|c|c|c|c|}
    \hline

    Time & Telescopes & Resolution & $\Delta\lambda$ &
    Seeing & Coherence Time & Air Mass & Phase\\

    \hline 
    \multicolumn{8}{|c|}{\textbf{Data set 1}}\\
    \hline 

    26/12/2004 3h48 & UT2-3-4 & 1500 & 2.04 - 2.08\,$\mu$m &
    0.53\,$\arcsec$ & 5.9\,ms & 1.293 & 0.36 \\

    26/12/2004 4h18 & UT2-3-4 & 1500 & 1.95 - 2.03\,$\mu$m &
    0.72\,$\arcsec$ & 4.5\,ms & 1.213 & 0.36 \\

    26/12/2004 4h35 & UT2-3-4 & 1500 & 2.02 - 2.10\,$\mu$m &
    0.60\,$\arcsec$ & 5.3\,ms & 1.180 & 0.36 \\

    26/12/2004 4h49 & UT2-3-4 & 1500 & 2.09 - 2.17\,$\mu$m &
    0.73 \,$\arcsec$ & 4.4\,ms & 1.158 & 0.36 \\

    \hline 
    \multicolumn{8}{|c|}{\textbf{Data set 2}}\\
    \hline 

    08/02/2006 3h42 & UT1-2-3 & 1500 & 1.64 - 1.79\,$\mu$m &
    1.06 \,$\arcsec$ & 4.5\,ms & 1.084 & 0.49 \\

    \hline 
    \multicolumn{8}{|c|}{\textbf{Data set 3}}\\
    \hline 

    30/12/2006 6h39 & UT1-3-4 & 35 & 1.64 - 1.79\,$\mu$m &
    0.82 \,$\arcsec$ & 8.8\,ms & 1.087 & 0.62 \\

    30/12/2006 6h39 & UT1-3-4 & 35 & 1.88 - 2.51\,$\mu$m &
    0.82 \,$\arcsec$ & 8.8\,ms & 1.087 & 0.62 \\

    \hline 
    \multicolumn{8}{|c|}{\textbf{Data set 4}}\\
    \hline 

    07/03/2007 3h07 & UT1-3-4 & 35 & 1.48 - 1.78\,$\mu$m &
    0.66 \,$\arcsec$ & 6.4\,ms & 1.123 & 0.51 \\

    07/03/2007 3h07 & UT1-3-4 & 35 & 1.88 - 2.38\,$\mu$m &
    0.66 \,$\arcsec$ & 6.4\,ms & 1.123 & 0.51 \\

    \hline 
    \multicolumn{8}{|c|}{\textbf{Data set 5}}\\
    \hline 

    31/03/2007 1h39 & UT1-2-4 & 1500 & 2.02 - 2.10\,$\mu$m &
    0.90 \,$\arcsec$ & 2.5\,ms & 1.129 & 0.77 \\

    31/03/2007 1h57 & UT1-2-4 & 1500 & 2.09 - 2.17\,$\mu$m &
    0.77 \,$\arcsec$ & 2.9\,ms & 1.154 & 0.77 \\

    \hline
  \end{tabular}
\end{table}

For all the observations, the Unit Telescopes were used. As explained
in Malbet et al.\cite{2007A&A...464...43M} and recalled in paper I,
the optical trains of the UT telescopes were affected by non-stationary
high-amplitude vibrations at the time of the observations. Therefore
the same data processing strategy as in paper I was used
for reducing the medium spectral resolution data (R=1500),
i.e. rejecting 80\% of the available data based on a SNR estimate on
the interferometric fringes.

For the low spectral resolution data, we used the new calibration
features developed in Millour et al. 2008\cite{Millour2008} to get the
most out of the new data. However, we did not use for this study the
coherence length and jitter corrections developed also in Millour et
al. 2008\cite{Millour2008} since they were not yet properly
tested. Therefore, we find that the low spectral resolution visibility
data are much less accurate than the medium spectral resolution
ones. At the other hand, the closure phase measurement in LR, has a
typical error of 0.3\deg\ i.e. $5\times10^{-3}$\,rad, which is highly
sufficient for this study. The old 2004 data-set was also reduced
using this new calibration strategy and we find very similar results
as in the paper I. A part of the data-set is shown in
Fig.~\ref{fig:allthedata}.

\subsection{Orbital solution refinement}
\label{subsection:observations}

The reference orbital solution used in this paper is the one computed
by North et al.\cite{2007MNRAS.377..415N}. We compared the previous
orbital solutions to the one we can find from a fit to the AMBER
interferometric data taking into account the synthetic spectra of the
two components, as explained in paper I. We find, using this new data,
a very good agreement with the published North et
al.\cite{2007MNRAS.377..415N} orbital elements. Our previous orbital
elements match this new purely interferometric estimate
marginally. This could be explained by the larger error coming from
the uncertainty of the spectropolarimetric measurements combined with
the sole astrometric point derived by AMBER in paper\,I.

\begin{figure}[htbp]
  \begin{center}
    \begin{tabular}{cccc}
      \includegraphics[width=7cm]{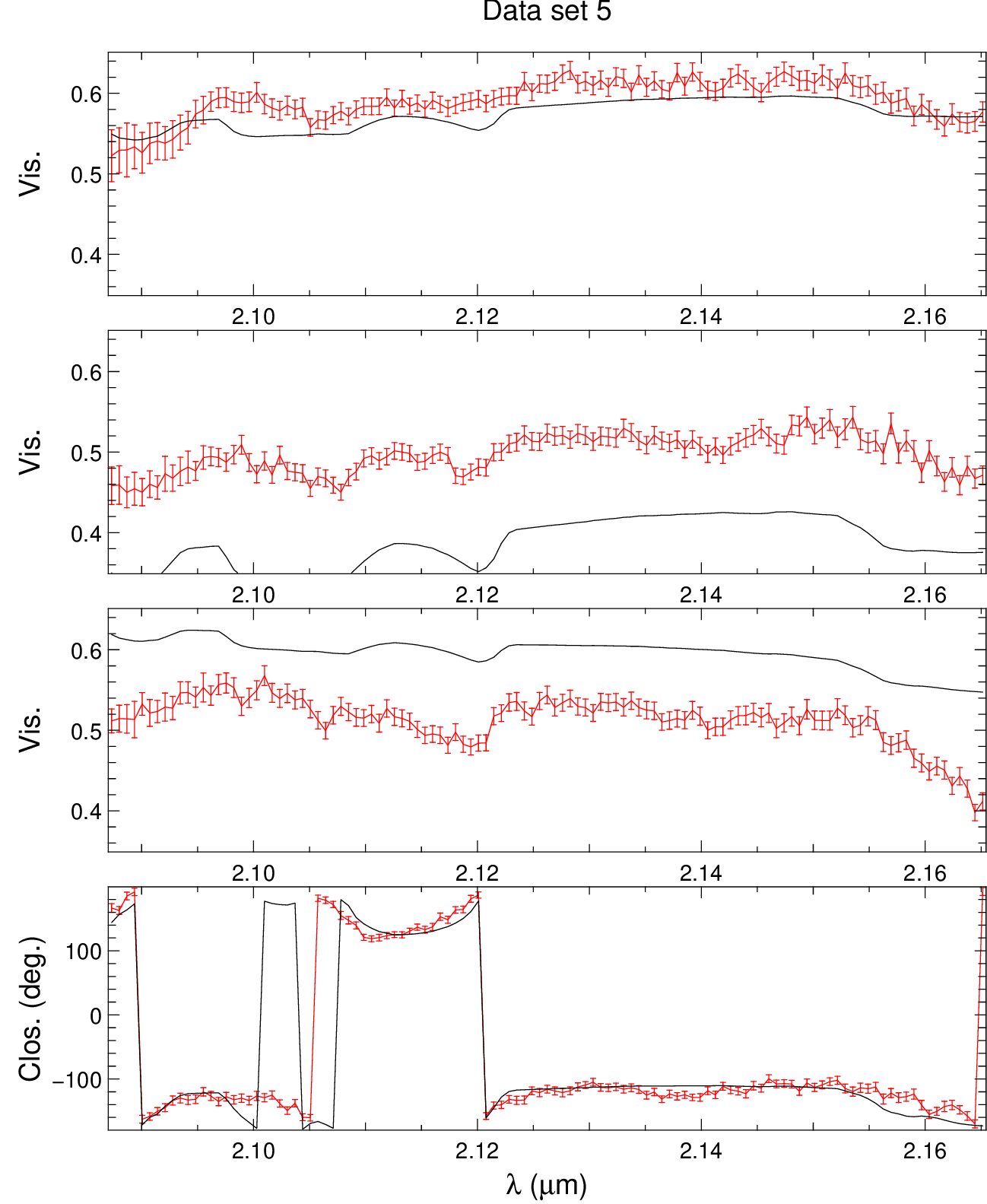}&
      \includegraphics[width=7cm]{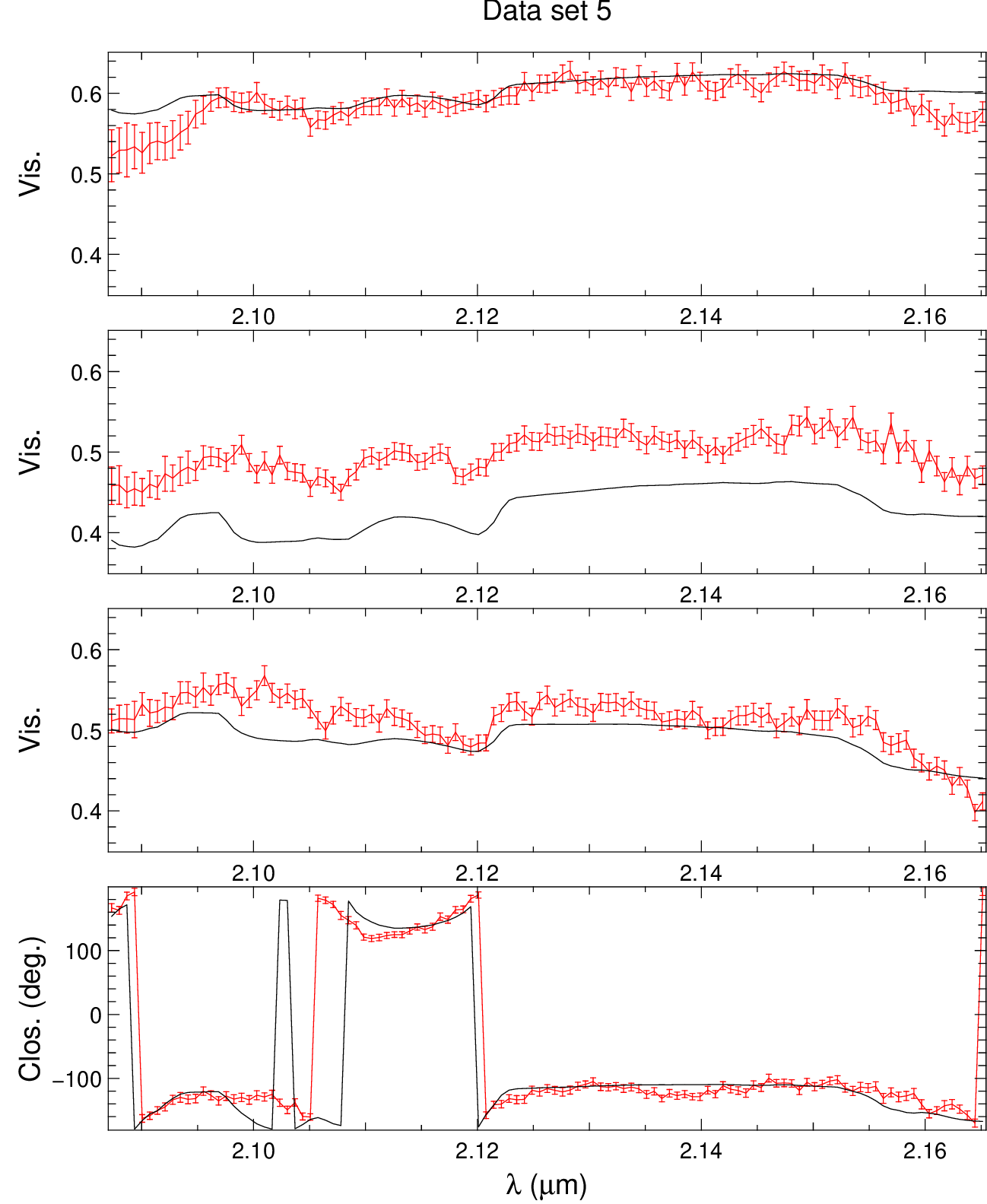}
    \end{tabular}
  \end{center}
  \caption
  % >>>> use \label inside caption to get Fig. number with \ref{}
  { \label{fig:allthedata} 
    A sub-sample of the data acquired on $\gamma^2$ Velorum in
    2006. The left panel shows the data together with a fit of the
    previously determined binary model. The right panel shows the same
    data with a fit including a 3rd component.
  }
\end{figure}

The resulting model can be seen in Fig.~\ref{fig:allthedata} as a thin black line,
to compare to the measurements plotted with error bars. The fit is
personally good but not formally good (reduced $\chi^2$ of about
2). Therefore one need to add, such as for the first AMBER paper, an
additional component to explain the discrepancy seen in the data.

\begin{table}[htbp]
  \centering
  \caption{
    \footnotesize{
      Orbital parameters from paper I, from North et
      al.\cite{2007MNRAS.377..415N} and from this work. Our AMBER data
      agrees very well to the published North et
      al.\cite{2007MNRAS.377..415N} ephemeris, but only marginally to
      our previous astrometric measurement.
    }
  }
  \label{obsLog}

  \begin{tabular}{|c|c|c|c|c|c|c|c|c|c|c|c|c|}
    \hline

    Reference & T$_0$ (days, fixed) & e & $\Omega$ (\deg) & $\omega$ (\deg) & i
    (\deg) & P (days, fixed) & a (mas)\\
    \hline
    Millour et al. A\&A 2007\cite{2007A&A...464..107M} & 50120.0 & 0.326 & 232.7 & 68.0 & 65.0 &
    78.53 & 4.8\\
    North et al. MNRAS 2007\cite{2007MNRAS.377..415N} & 50120.4 & 0.334 & 247.7 & 67.4 & 65.5 & 78.53
    & 3.57\\
    This work & 50120.4 & 0.334 & 253 & 67 & 64 &  78.53 & 3.56\\

    \hline
  \end{tabular}
\end{table}

\subsection{A third light source in the system}
\label{subsection:observations}

To investigate the presence of an additional contribution of the wind-wind
collision to the overall near-IR continuum flux of the system, we added a third
component to our 2-stars model, which is spatially unresolved and whose position
is restricted within the line joining the two stars. We also
restricted the spectrum of this additional component to be a fraction
of the total spectrum (i.e. with a pseudo-achromatic contribution). Therefore, the corresponding model can be easily described as a 3-body model, the binary star being fixed by
the previous study and the 3rd component being described by only 2
parameters: $R$, the flux fraction of this additional component, and
$r$, the relative coordinate of the source between the WR star and the
O star. In this way, $R$ and $r$ are therefore 2 numbers between 0 and 1.

% 
%%%%%%%%%%%%%%%%%%%%%%%%%%%%%%%%%%%%%%%%%%%%%%%%%%%%%%%%%%%%%%%%%%%%%%% 
% 

Given this simple model, one can vary $R$ and $r$ and compute a
$\chi^2$ map. The
striking fact is that the minimum of $\chi^2$ does not match to a flux
$R$ for the third component of 0, but to a fraction of light of
5-10\%, where the $\chi^2$ goes from 2 in the previous study to
1.6 here. This fact is probably confirming the speculation on paper\,I
of an additional light source contributing to about 5\% of the overall
flux. Moreover, we are able to constrain the position of this
additional source, which is roughly centered at 2/3 in-between the WR
and O stars.

% This finding has strong implications to the wind momentum ratio, and
% it seems that the WR wind momentum is only 2 times stronger than the O
% wind%  ({\bf PHRASE D'EXPLICATION+REFERENCE NECESSAIRE.}
% . Comparison to what would be expected is under progress.

% \begin{figure}[htbp]
%   \begin{center}
%     \includegraphics[width=17cm,angle=0]{double.eps}
%   \end{center}
%   \caption
%   %   >>>> use \label inside caption to get Fig. number with \ref{}
%   { \label{fig:chi2Maps}
%     $\chi^2$ maps with $R$ representing the fraction of light of
%     the third source to the overall flux and $r$ the fraction of the
%     linear coordinate between the WR and O stars (i.e. $r=0$ means the
%     source is superposed to the WR star and $r=1$ means it is
%     superposed to the O star). The right map is just a zoom into a
%     fraction of the 1st map.
%   }
% \end{figure} 

\section{Dusty Wolf-Rayet stars}
\label{section:intro}

A challenging problem is to understand how dust can form in the
hostile  environment of hot Wolf-Rayet stars. We successfully observed
so far with AMBER 2 dusty Wolf-Rayet stars, which illustrate the
potential of interferometry to contribute to answer to the fascinating
question of binarity of these targets.

\subsection{WR48a(-b)}

\begin{figure}[htbp]
  \begin{center}
    \begin{tabular}{cccc}
      \includegraphics[width=7cm]{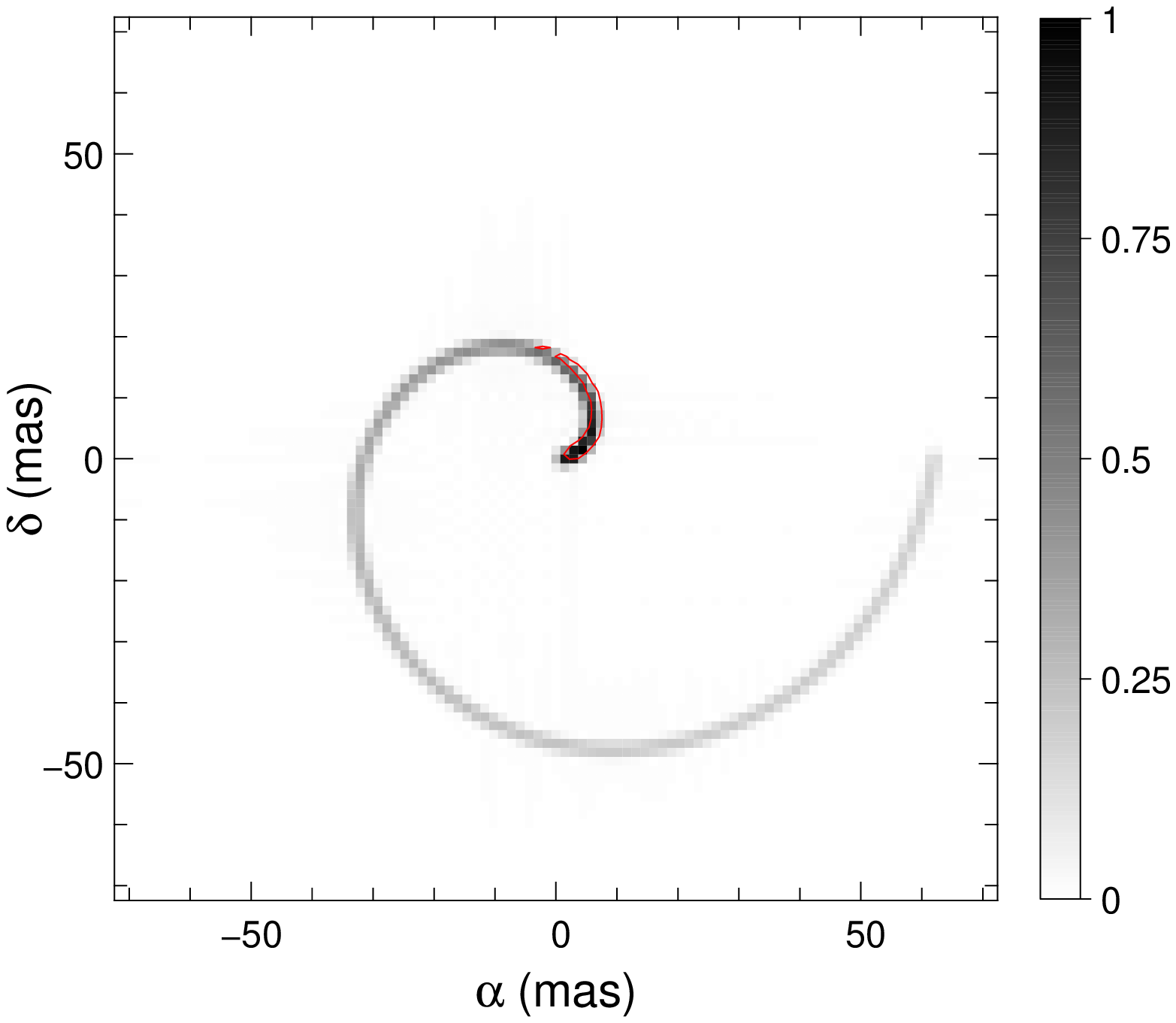}&
      \includegraphics[width=8cm]{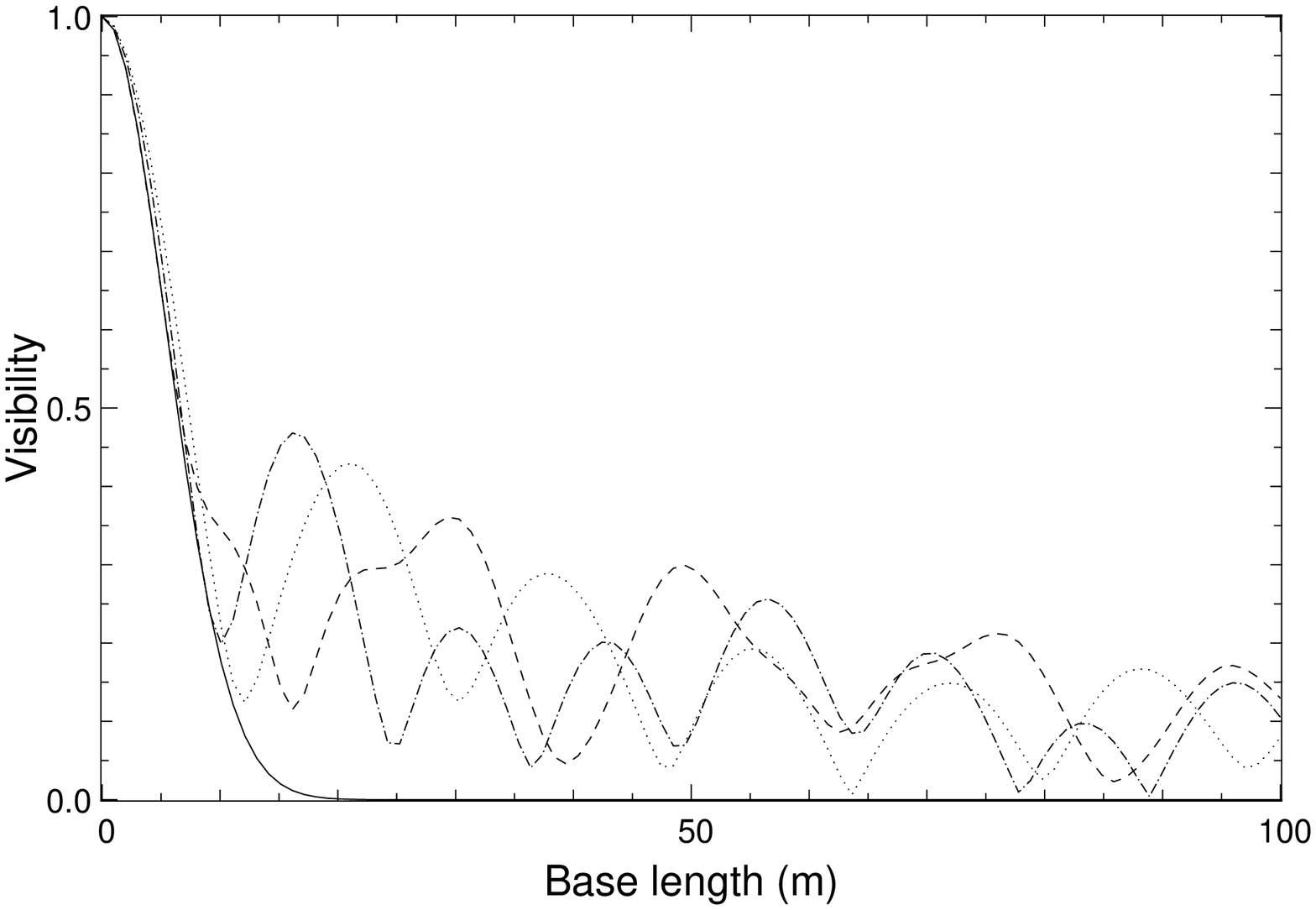}\\
      \includegraphics[width=8cm,angle=-360, origin=c]{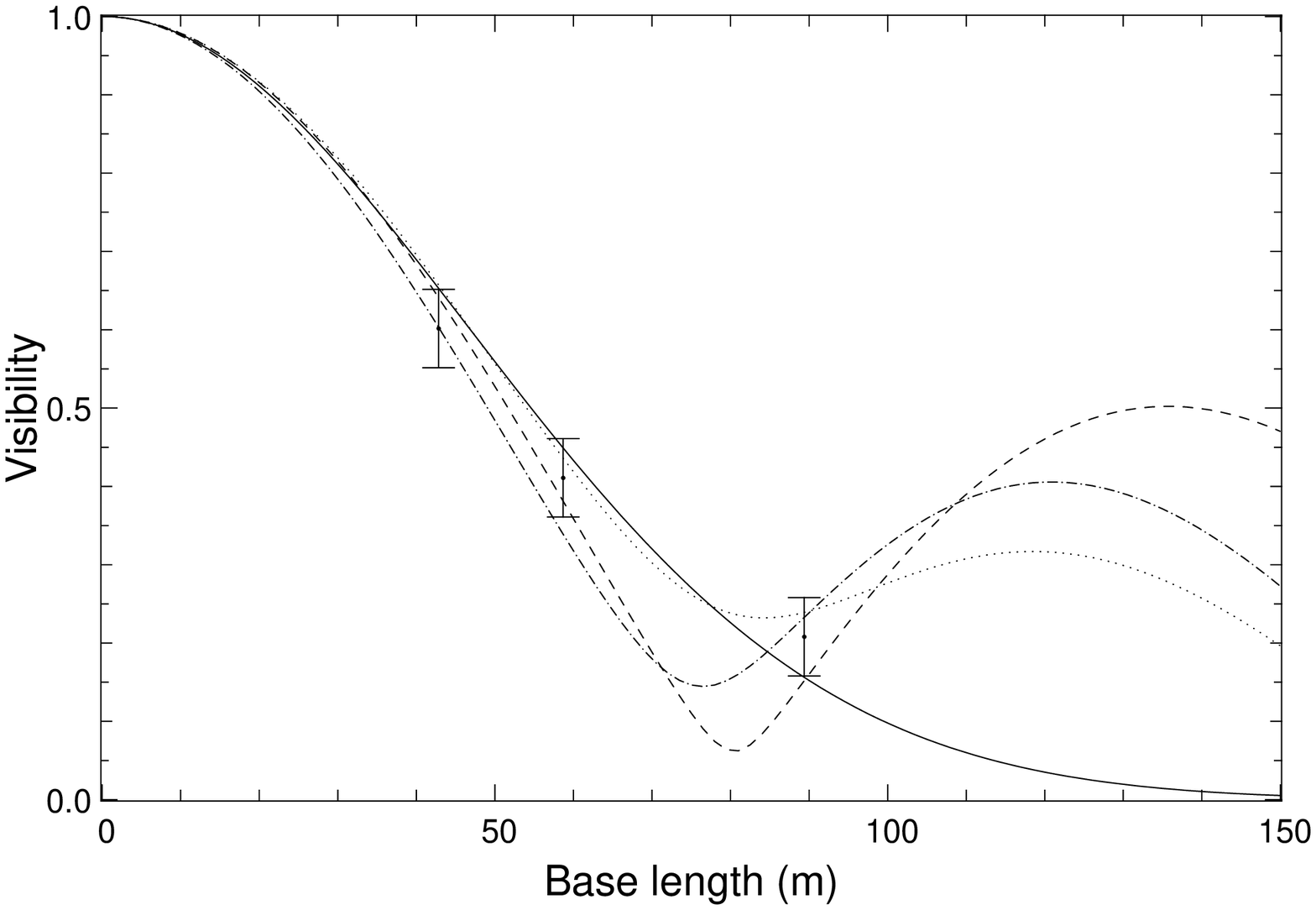}&
      \includegraphics[height=8cm,angle=-90, origin=c]{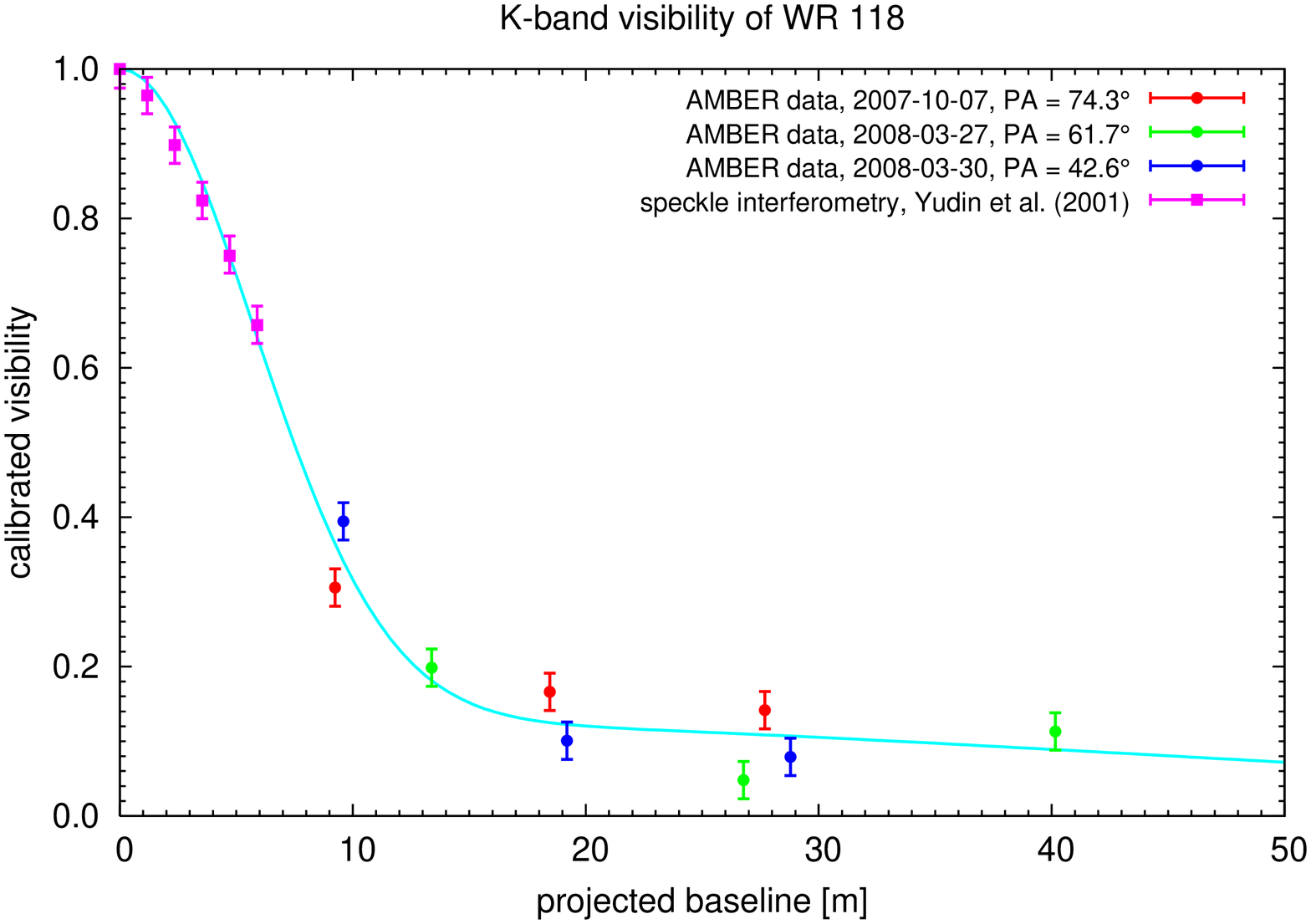}
    \end{tabular}
  \end{center}
  \caption
  % >>>> use \label inside caption to get Fig. number with \ref{}
  { \label{fig:pinwheelModels} 
    \emph{Top-left}: A phenomenological model for a \emph{pinwheel} dusty
    nebula. The tail is 10 times fainter than the core of the nebula.
    \emph{Top-right}: Visibility cuts at 3 different hour angles (dashed
    lines) for the previous model, compared to a Gaussian model with
    the same typical size (full line).
    \emph{Bottom-left}: The WR48a data compared to 2 different models: a
    Gaussian source (full line) and a pinwheel at three different
    position angles (dashed lines). The visibilities marginally fit to
    a Gaussian model, and could be better explained by the presence of
    a pinwheel nebula.
    \emph{Bottom-right}: 
    $K$-band visibility of the carbon-rich Wolf-Rayet star WR~118
    obtained with bispectrum speckle interferometry (magenta bullets;
    Yudin et al.\ 2001) and with VLTI/AMBER (three measurements;
    red/green/blue bullets). In addition to the extended dust envelope,
    the VLTI/AMBER measurements reveal a compact component contributing
    $\sim15\%$ to the total $K$-band flux. To guide the eye and to give
    a rough idea of the spatial scales involved, the light blue curve
    shows the fit of a double-Gaussian intensity profile, revealing
    components of $\sim30$ and $\sim3$~mas. A comparison with
    the top-right figure shows that the visibility signature of
    WR~118 found from the speckle + AMBER observations could be
    qualitatively explained by a spiral-like dust distribution. This
    would support the idea that WR~118 is a close binary system where
    the dust production takes place in a wind-wind collision zone,
    similar to the well-known examples WR~98a and WR~104.
  }
\end{figure}

WR48a is a carbon-rich Wolf-Rayet stars (WC) producing dust in the
form of large or mini eruptions, that suggest the presence of a
companion. It is found to be within $\sim$1' of the two heavily
reddened, optically visible clusters Danks 1 and 2 – which are
themselves separated by only $\sim$2'\cite{2004A&A...427..839C}. WR
48a is present and visible as the bright object in the
cleared centre of the field of the Mid-Course experiment observations
(MSX6C G305.3614+00.0561) and a close object that seems to be also a
dusty WR star (MSX6C G305.4013+00.0170) was found close to it, that we
hereafter denote as WR48a-b. No optical counterpart is reported in the
literature and the JHK photometry for this source from the 2MASS
survey show an emission falloff indicative of an object with high
extinction (AV$\geq$10 mag). With fluxes of 53.2, 54.7 and 42.0 Jy for
bands A, C, D respectively (corresponding to N band), WR48a-b is
almost as bright as WR48a in the mid-IR. The discovery of a WR is
always an important event since these objects are very rare (no more
than 500-1000 in our galaxy) and it turns out that most of the dusty
WR may harbor pinwheel nebulae as shown recently by the observations
of Tuthill et al.\cite{2006Sci...313..935T}. WR48a was successfully
observed in P78 (078.D-0503D) by the AMBER/VLTI instrument. The size
measured is quite large and these two objects are probably two good
candidates to harbor 'pinwheel' nebulae. WR48a-b was also successfully
observed by AMBER with a 43, 59 and 89m baselines in the low
resolution mode: the visibilities are quite low, between 0.2 and 0.5
(see Fig.~\ref{fig:pinwheelModels}), implying that the object has an
angular diameter of about 4-6mas in K band, and that mid-IR sizes
might be of the order of 10-15mas. Definitely these two targets are
interesting, because they produce dust and hence are likely binary
candidate, and also because they belong to a young star forming region
whose distance is more accurately known that usual.

% 
%%%%%%%%%%%%%%%%%%%%%%%%%%%%%%%%%%%%%%%%%%%%%%%%%%%%%%%%%%%%%%%%%%%%%%% 
% 

\subsection{WR118}

WR~118 is a highly evolved, carbon-rich Wolf-Rayet star of spectral
type WC10. It is the third brightest Wolf-Rayet star in the $K$ band
($K=3.65^{\rm m}$), and the large IR-excess is usually attributed to an
envelope composed of carbonaceous dust. Since no remarkable changes in
the dust emission have been observed in the past two decades, WR~118
is classified as permanent dust maker. The extended dust envelope of
WR~118 was successfully resolved, for the first time, by Yudin et
al.\cite{2001A&A...379..229Y} using bispectrum speckle interferometry
with the BTA 6~m telescope. They found that up to the cutoff-frequency
the $K$-band visibility is spherically symmetric and drops to 0.66.
From their 1D-radiative transfer modeling of the $K$-band visibility
and spectral energy distribution Yudin et al.\ concluded that the
apparent diameter of WR~118's inner dust shell boundary is
$17\pm1$~mas. Furthermore, they found that the SED and
$K$-band visibility can either be reproduced with a two-component
grain model containing large (grain size $a=0.38\,\mu$m) as well as
small grains ($a=0.05\,\mu$m), or with a model where the grain sizes
follow a $n(a) \sim a^{-3}$ size distribution. At the inner edge of
the dust shell, the dust temperature and density of the best-fitting
model are $1750\pm100$~K and $(1.0\pm0.2)\cdot10^{19}\,{\rm g/cm}^{3}$.

Recently, the first $K$-band measurements of WR~118 have been obtained
with VLTI/AMBER in low-spectral resolution mode. For all three
observations,a linear baseline configuration was used, with projected
baselines ranging from $\sim10$ to
$\sim$40~m. Fig.~\ref{fig:pinwheelModels} 
presents the AMBER visibilities at $2.2\,\mu$m together with the
azimuthally averaged $K$-band visibility from the
speckle-interferometric measurements of Yudin et
al.\cite{2001A&A...379..229Y}. The figure illustrates that the AMBER
measurements nicely supplement the previous speckle
observations. Moreover, these data show some interesting results:
\begin{itemize}
\item At first glance, the AMBER visibilities suggest that there is an
  unresolved component contributing approximately 15\% to the total
  $K$-band flux.
\item Taking a closer look at the AMBER data, one can see
  that the $K$-band visibility is not spherically symmetric within the
  error bars (see, for instance, the red and blue data points) and the
  visibility at position angle $61.7^\circ$ is not monotonically
  decreasing.
\item On the other hand, the comparison of the AMBER and speckle
  measurements with the model visibility in
  Fig.~\ref{fig:pinwheelModels} 
  illustrates that the overall shape of the measured visibility of
  WR~118, i.e.\ the noticeable, basically spherically-symmetric drop  at
  short baselines plus the plateau-like, non-spherically-symmetric trend
  at baselines larger than 10~m, is in qualitative agreement with the
  visibility signature expected from a spiral-like dust
  distribution.
\end{itemize}
All these clues might imply that WR~118 harbours a binary system
where circumstellar dust is produced in the co-rotating wind-wind
collision zone where the physical conditions are favourable for dust
production. Currently, the interpretation of the recently obtained
AMBER data is subject to a more detailed analysis.

% 2FEAA0A188
% 881A0AAEF2
%%%%%%%%%%%%%%%%%%%%%%%%%%%%%%%%%%%%%%%%%%%%%%%%%%%%%%%%%%%%% 
%%%%% References %%%%%

\section{Conclusion}

We presented here a collection of AMBER observations on several
Wolf-Rayet stars, and the different phenomena that can be
characterized using the VLTI today:
\begin{itemize}
\item the wavelength-dependent spatial extent of the optically thick
  wind of single Wolf-Rayet stars (WR79a),
\item the presence of wind-collision zones by means of the free-free
  emission of the compressed gas,
\item the detection of ``pinwheel'' nebulae in dusty Wolf-Rayet stars
  which are too far to be resolved by single-dish telescopes.
\end{itemize}

The corresponding data processing and interpretations are still under
progress, but they all show the great potential of AMBER to resolve
the many different geometric features that evolved stars such as
Wolf-Rayet can show to us.

\bibliographystyle{spiebib}   %>>>> makes bibtex use spiebib.bst
\bibliography{biblio2}   %>>>> bibliography data in report.bib

\end{document}